\documentclass[aps,preprint]{revtex4}%
\usepackage{amsfonts}
\usepackage{amsmath}
\usepackage{amssymb}
\usepackage{graphicx}
\usepackage{natbib}%
\usepackage{xr}
\usepackage[caption=false]{subfig}
\providecommand{\U}[1]{\protect\rule{.1in}{.1in}}
\begin{document}
\preprint{ }
\title{Adsorption of highly charged Gaussian polyelectrolytes to oppositely charged surfaces}

\author{Sandipan Dutta}
\affiliation{ Asia Pacific Center for Theoretical Physics, Pohang, Gyeongbuk, 790-784, Korea}
\author{and Y.S. Jho}
\email{ysjho@apctp.org}
\affiliation{Department of Physics, Pohang University of Science and Technology, Asia Pacific Center for Theoretical Physics, Pohang, Gyeongbuk, 790-784, Korea}

\begin{abstract}
In many biological processes highly charged biomolecules are adsorbed into oppositely charged surfaces of macroions and membranes. They form 
strongly correlated structures close to the surface which can not be explained by the conventional Poisson-Boltzmann theory. Many of the
flexible biomolecules can be described by Gaussian polymers. In this work strong coupling theory is used to study the adsorption of highly charged Gaussian
polyelectrolytes. Two cases of adsorptions are considered, when the Gaussian polyelectrolytes are confined a) by one charged wall, and
b) between two charged walls. The effects of salt and the geometry of the polymers on their adsorption-depletion transitions in the strong 
coupling regime are discussed. 
\end{abstract}
%\keywords{}

\maketitle

\section{Introduction}
\label{Sec0}
The phenomenon of polymer adsorption is of great interest in biological sciences as it is the key mechanism behind many of the biological processes
like DNA winding around nucleosome cores \cite{luger1997crystal,cherstvy2009positively}, binding of genomes in RNA viruses by capsid proteins
\cite{belyi2006electrostatic,forrey2009electrostatics}, tau association to microtubules \cite{jho2010monte} and polyelectrolyte multilayer vesicles \cite{caruso1999investigation}.
Of particular 
interest is that of adsorption caused by electrostatic forces as many biomolecules like DNA and RNA are highly charged and they frequently
bind themselves to oppositely charged protein or membrane surfaces. There is an extensive literature both in theory \cite{
baumgartner1991effects,von1994adsorption,linse1996adsorption,shafir2003adsorption,winkler2006critical,man2008adsorption,
livcer2010polyelectrolyte,wang2011charge,cherstvy2012polyelectrolyte,de1979scaling,podgornik1990forces} as well as in simulations
\cite{chodanowski2001polyelectrolyte,kong1998monte,welch2000dendrimer,messina2004polyelectrolyte,ulrich2005complexation,forsman2012polyelectrolyte,xie2013polyelectrolyte,
stoll2002polyelectrolyte,källrot2010dynamics,ulrich2006many,cao2011interaction,carnal2011adsorption,cerda2009understanding,jeong2014monte} on electrostatic driven adsorption of 
flexible polyelectrolyte (PE) chains onto surfaces. Experimental studies have also been performed to investigate the dependence of the critical adsorption
threshold on the PE charges, salt concentration and the surface charge density of the fixed charge distribution \cite{mcquigg1992critical,
feng2001critical,kayitmazer2005role,cherstvy2011polyelectrolyte}. A PE is adsorbed into a surface when the electrostatic attractions between the PE and the surface
overcomes the entropy of the PE chain. The presence of salt plays an important role in the adsorption process as higher salt concentration screens 
the electrostatic attractions thus inhibiting adsorption.

Several theoretical models exist to describe PE adsorption into charged surfaces. Most of these models use mean field description \cite{chatellier1996adsorption,
joanny1999polyelectrolyte,wiegel1977adsorption,muthukumar1987adsorption}, various scaling theories \cite{borukhov1998scaling,borukhov1995polyelectrolyte,
borisov1994polyelectrolyte,dobrynin2001adsorption,dobrynin2000adsorption,dobrynin2005theory} or other phenomenological criteria for adsorption \cite{netz1999adsorption,
netz2003neutral}. Wiegel \cite{wiegel1977adsorption} was one of the first to obtain a closed expression for the adsorption threshold for
Gaussian polyelectrolytes (GPEs) 
in the mean field approximation using polymer field theory. Muthukumar \cite{muthukumar1987adsorption} later extended this work to include excluded 
volume interactions among PE chains.
In references \cite{shafir2003adsorption,borukhov1998scaling,borukhov1995polyelectrolyte} a coupled Poisson – Boltzmann and Edwards polymer
density equations \cite{edwards1965statistical} have been used to obtain scaling laws for adsorption-depletion transitions for PEs.
Dobrynin \textit{et al.} \cite{dobrynin2000adsorption,dobrynin2001adsorption} proposed a Wigner liquid structure for the adsorbed polyelectrolyte layer
based on the idea of counterion condensation by Shklovskii \cite{shklovskii1999wigner,shklovskii1999screening,perel1999screening,RevModPhys.74.329}. 
The conventional Poisson-Boltzmann formalism fails inside the Wigner layer and strong coupling (SC) theories \cite{Netz,moreira2001binding,
shklovskii1999wigner,kanduvc2009role,naji2013perspective,levin2002electrostatic,naji2005counterions,boroudjerdi2005statics,RevModPhys.74.329}
are more appropriate descriptions in these domains. In fact many highly charged biomolecules act as multivalent counterions with strong 
many-body correlations and the strong coupling theories, which were developed to explain the like-charged attractions in 
multivalent counterions (for which the Coulomb coupling is large $\Xi >> 1$ as opposed to the mean field approximation
which is valid in the weak coupling $\Xi < 1$), provide a better description of their thermodynamics. 
The SC theories however were originally developed for point particles which is not applicable to these biomolecules because 
of their finite geometries. Many of the flexible biomolecules would be better approximated as Gaussian polymers.
Recently the SC theories have been 
extended to rodlike polyelectrolytes \cite{bohinc2012interactions, kim2008attractions,strongrods,RevModPhys.74.329}. But their 
application to the GPEs specially for the adsorption phenomena have not been done yet, to the best of the authors' knowledge.

In the current work we develop a theoretical formalism to study the adsorption of GPEs into oppositely charged surfaces in the
SC regime. This formalism is closely based on an earlier work by the current authors \cite{strongrods} and is inspired by the strong coupling theory 
developed by Netz and his co-workers \cite{Netz, moreira2001binding}. The outline of the paper is as follows. In Section \ref{Sec1} we obtain an expression 
for the partition function in the SC regime and from that 
the density of PEs near the surface. In SC the thermodynamic parameters can be expanded in $1/\Xi$ and at very strong coupling only the lowest order terms
dominates. In a recent formalism based on Wigner crystal, \v{S}amaj and Trizac \cite{samajPRE, samajPRL} obtained the leading order correction to be 
$1/\sqrt{\Xi}$ not $1/\Xi$ as in Netzs' theory \cite{hatlo2010electrostatic}. However the zeroth order term of their densities agrees with that of Netz.
Therefore we use the zeroth order term of the density distribution to study the adsorption of GPEs in this work. We consider the 
confinement by both one oppositely charged wall and two charged walls in the salt-free regime in Section \ref{Sec2}. 
The walls are considered to be impenetrable in that the density of PEs vanishes on and inside the
walls similar to the condition imposed in References \cite{wiegel1977adsorption,muthukumar1987adsorption,shafir2003adsorption}. In the absence of salt 
whereas we always get adsorption in the SC regime in case of one wall confinement, for two walls we get depletion. In Section \ref{Sec3} we consider salt
implicitly through a screening potential and analyze the dependence of the adsorption-depletion transition threshold on the 
polymer geometry and the salt concentration. The results are summarized and future extensions of this work is discussed in Section \ref{Sec4}.

\section{Thermodynamics of Gaussian Polyelectrolytes in the SC limit}
\label{Sec1}
We consider a system of $N$ GPE counterions in presence of a fixed charge distribution $-\rho_{f}(\mathbf{x})$.
The fixed charged distribution has a surface charge density $-\sigma_s$. In addition there is an external potential $-h(\mathbf{x})$. 
Overall the system is charge neutral. Each polymer is made of $L$ monomers each of length $b$ and charge $q$ distributed uniformly over the monomer.
In this work the notation $L$ would be used interchangeably for both the number of monomers in each polymer and the length of the polymers $Lb$.
The electrostatic interaction is denoted by $v(\mathbf{x})$. The position coordinate at a segment $s$ of the polymers is
parameterized by a field $\mathbf{r}(s)$ and the segment density function of the polymer 
counterions by
\begin{equation}
 \hat{\rho}(\mathbf{x}) = \sum\limits_{i=1}^{N}\int_{0}^{L}ds\delta(\mathbf{r}_{i}(s)-\mathbf{x}),
 \label{eq1.1}
\end{equation}
where $\mathbf{r}_{i}(s)$ denotes the polymer field for the $i$th polymer. The Hamiltonian of the system is given by
\begin{align}
 \beta \mathcal{H}_N & = \sum_{i=1}^N\beta \mathfrak{h}_0^i +\frac{l_B}{2}\int d\mathbf{x}d\mathbf{x}^{\prime}\left[q\hat{\rho}(\mathbf{x})-\rho_{f}(\mathbf{x})
 \right]v(\mathbf{x}-\mathbf{x}^{\prime}) 
 \left[q\hat{\rho}(\mathbf{x}^{\prime})-\rho_{f}(\mathbf{x}^{\prime})\right] -\frac{l_B}{2}q^2N\beta v_s \nonumber\\ 
  & - \int d\mathbf{x}h(\mathbf{x})\hat{\rho}(\mathbf{x}),
  \label{eq1.2}
\end{align}
where $\mathfrak{h}_0^i$ is the single Gaussian polymer Hamiltonian, $\beta = 1/k_BT$ inverse temperature and $l_B = \beta e^2/\epsilon$ is the Bjerrum length.
We use the dielectric constant of $\epsilon = 80$ in the rest of the discussions. The intra-molecular energy due the connectivity of 
the Gaussian polymers is given by  
\cite{fredrickson2006equilibrium}
\begin{equation}
\beta\mathfrak{h}_0 = \frac{3}{2b^2}\int_0^L ds\biggl\vert\frac{d\mathbf{r}(s)}{ds}\biggr\vert^2.
 \label{eq1.3}
\end{equation}
$v_s$ is the electrostatic self interaction between the different segments of the same polymer,
\begin{equation}
v_s = \int_0^L ds \int_0^L ds^{\prime}v(\vert\mathbf{r}_i(s)-\mathbf{r}_i(s^{\prime})\vert).
 \label{eq1.4}
\end{equation} 

We convert all the quantities above to dimensionless form. The position coordinates $\mathbf{x}$ 
and the polymer fields $\mathbf{r}$ are scaled by a Gouy-Chapman-like length scale $\mu = 1/(2\pi ql_B\sigma_s)$ , 
$\widetilde{\mathbf{x}}=\mathbf{x}/\mu$ and  $\widetilde{\mathbf{r}}(s)=\mathbf{r}(s)/\mu$.
Following Netz we introduce a dimensionless coupling parameter $\Xi = 2\pi q^3l_B^2\sigma_s$.
Similarly the dimensionless fixed charge distribution is defined by $\widetilde{\rho_{f}}(\mathbf{x}/\mu)
= \mu\rho_f(\mathbf{x})/\sigma_s$ and the dimensionless polymer density by $\hat{\rho}(\mathbf{x}/\mu) = \mu^3\hat{\rho}(\mathbf{x})$. 
In dimensionless form the Hamiltonian in equation \eqref{eq1.2} becomes
\begin{align}
 \widetilde{\mathcal{H}}_N & = \sum_{i=1}^N\widetilde{\mathfrak{h}}_0^i +\frac{\Xi}{2}\int d\widetilde{\mathbf{x}}d\widetilde{\mathbf{x}}^{\prime}
 \hat{\rho}(\widetilde{\mathbf{x}})\widetilde{v}(\vert\widetilde{\mathbf{x}}-\widetilde{\mathbf{x}}^{\prime}\vert)\hat{\rho}(\widetilde{\mathbf{x}}^{\prime})
 +\frac{1}{8\pi^2\Xi}\int d\widetilde{\mathbf{x}}d\widetilde{\mathbf{x}}^{\prime} 
 \widetilde{\rho}_f(\widetilde{\mathbf{x}})\widetilde{v}(\vert\widetilde{\mathbf{x}}-\widetilde{\mathbf{x}}^{\prime}\vert)\widetilde{\rho}_f(\widetilde{\mathbf{x}}^{\prime})
 \nonumber \\& + \int d\widetilde{\mathbf{x}}\widetilde{u}(\widetilde{\mathbf{x}})\hat{\rho}(\widetilde{\mathbf{x}})
  -\frac{\Xi}{2}N\widetilde{v}_s - \int d\widetilde{\mathbf{x}}h(\widetilde{\mathbf{x}})\hat{\rho}(\widetilde{\mathbf{x}}).
  \label{eq1.5}
\end{align}
The fourth term on the RHS of the above equation is the external potential on the GPEs due to the fixed charge distribution  
\begin{equation}
\widetilde{u}(\widetilde{\mathbf{x}}) = -\frac{1}{2\pi}\int d\widetilde{\mathbf{x}}^{\prime}\widetilde{\rho}_f(\widetilde{\mathbf{x}}^{\prime})
\widetilde{v}(\vert\widetilde{\mathbf{x}}-\widetilde{\mathbf{x}}^{\prime}\vert).
 \label{eq1.6}
\end{equation}
We call it the wall potential. In many cases $\widetilde{u}(\widetilde{\mathbf{x}})$ diverges due to the long range nature of the Coulomb interactions. 
However the divergence is removed if a vanishing term \cite{Netz}
\begin{equation}
\frac{1}{4\pi^2\Xi}\int d\widetilde{\mathbf{x}}d\widetilde{\mathbf{x}}^{\prime} 
 \widetilde{\rho}_f(\widetilde{\mathbf{x}})\widetilde{v}(\vert\widetilde{\mathbf{x}}-\widetilde{\mathbf{x}}_0\vert)\widetilde{\rho}_f(\widetilde{\mathbf{x}}^{\prime})
 -\frac{N}{2\pi}\int d\widetilde{\mathbf{x}}^{\prime}\widetilde{\rho}_f(\widetilde{\mathbf{x}}^{\prime})\widetilde{v}(\vert\widetilde{\mathbf{x}}-\widetilde{\mathbf{x}}_0\vert) = 0
 \label{eq1.7}
\end{equation}
is added to the Hamiltonian and the wall potential is redefined as 
\begin{equation}
\widetilde{u}(\widetilde{\mathbf{x}}) = -\frac{1}{2\pi}\int d\widetilde{\mathbf{x}}^{\prime}\widetilde{\rho}_f(\widetilde{\mathbf{x}}^{\prime})
\biggl[\widetilde{v}(\vert\widetilde{\mathbf{x}}-\widetilde{\mathbf{x}}^{\prime}\vert)- \widetilde{v}(\vert\widetilde{\mathbf{x}}^{\prime}
-\widetilde{\mathbf{x}}_0\vert)\biggr].
 \label{eq1.8}
\end{equation}
Note that this added term vanishes due to the charge neutrality condition $\int d\widetilde{\mathbf{x}}\widetilde{\rho}_f(\widetilde{\mathbf{x}}) =
2\pi\Xi N $. The coordinate point $\widetilde{\mathbf{x}}_0$ in the above equation is chosen appropriately to cancel out the divergence in 
$\widetilde{u}(\widetilde{\mathbf{x}})$. After this modification the Hamiltonian reads
\begin{align}
 \widetilde{\mathcal{H}}_N & = \sum_{i=1}^N\widetilde{\mathfrak{h}}_0^i +\frac{\Xi}{2}\int d\widetilde{\mathbf{x}}d\widetilde{\mathbf{x}}^{\prime}
 \hat{\rho}(\widetilde{\mathbf{x}})\hat{\rho}(\widetilde{\mathbf{x}}^{\prime})\widetilde{v}(\vert\widetilde{\mathbf{x}}-\widetilde{\mathbf{x}}^{\prime}\vert)
 +\frac{1}{4\pi^2\Xi}\int d\widetilde{\mathbf{x}}d\widetilde{\mathbf{x}}^{\prime} 
 \widetilde{\rho}_f(\widetilde{\mathbf{x}})\widetilde{\rho}_f(\widetilde{\mathbf{x}}^{\prime})\biggl[\widetilde{v}(\vert\widetilde{\mathbf{x}}-\widetilde{\mathbf{x}}^{\prime}\vert)/2
 \nonumber \\& -\widetilde{v}(\vert\widetilde{\mathbf{x}}-\widetilde{\mathbf{x}}_0\vert)\biggr] + \int d\widetilde{\mathbf{x}}\widetilde{u}(\widetilde{\mathbf{x}})\hat{\rho}(\widetilde{\mathbf{x}})
  -\frac{\Xi}{2}Nv_s - \int d\widetilde{\mathbf{x}}h(\widetilde{\mathbf{x}})\hat{\rho}(\widetilde{\mathbf{x}}).
  \label{eq1.9}
\end{align}

The grand partition function of the system is given by
\begin{equation}
  Z = \sum\limits_{N=0}^{\infty}\frac{\lambda^N\mu^{3N}}{N!\lambda_T^{3N}}\int\left[\prod\limits_{i=1}^{N}\mathcal{D}\widetilde{\mathbf{r}}_i
  \widetilde{\Omega}(\widetilde{\mathbf{r}}_i)\right]\exp(-\widetilde{\mathcal{H}}_N[\{\widetilde{\mathbf{r}}_i\}]),
  \label{eq1.10}
\end{equation}
where $\lambda$ denotes the fugacity, $\lambda_T = \sqrt{h^2\beta/2\pi m}$ the thermal wavelength and $\mathcal{D}$ the integral over the polymer field 
configurations. The square brackets represents the functional dependence of $\widetilde{H}_N$ on the polymer fields $\widetilde{\mathbf{r}}(s)$. 
The term $\mu^{3N}$ comes from the rescaling of the polymer fields $\widetilde{\mathbf{r}}_i$. The function $\widetilde{\Omega}(\widetilde{\mathbf{x}})$
represents the amount of available space
to the counterions which might be due to the confinement by hard walls or traps. Since the SC theory is exact in the limit when $1/\Xi\rightarrow 0$,
we introduce a scaled fugacity defined by $\Lambda =2\pi\mu^3\lambda\Xi/\lambda_T^3$ that enable us to rewrite equation \eqref{eq1.10} in
a series in $1/\Xi$
\begin{equation}
  Z = \sum\limits_{N=0}^{\infty}\left(\frac{\Lambda}{2\pi\Xi}\right)^NQ_N.
  \label{eq1.11}
\end{equation}
$Q_N$ is the canonical partition functions for $N$ polymers and has the functional form
\begin{align}
Q_N[h-\widetilde{u},\widetilde{v}]  & = \frac{Z_0}{N!}\int\left[\prod\limits_{i=1}^{N}\mathcal{D}\widetilde{\mathbf{r}}_i
  \widetilde{\Omega}(\widetilde{\mathbf{r}}_i)\right]\exp\biggl(-\sum\limits_{i=1}^NH_0[\widetilde{\mathbf{r}_i}]-\Xi\sum_{i<j}\int_{0}^{N}ds
  \int_{0}^{N}ds^{\prime}\widetilde{v}(\vert\widetilde{\mathbf{r}_i}(s)- \widetilde{\mathbf{r}_j}(s^{\prime})\vert) \nonumber\\ & - \sum\limits_{i=1}^N\int_0^Nds
  \widetilde{u}[\widetilde{\mathbf{r}_i}(s)]+ \sum\limits_{i=1}^N\int_0^Ndsh[\widetilde{\mathbf{r}_i}(s)]\biggr),
 \label{eq1.12}
\end{align}
where 
\begin{align}
Z_0  = \exp\left(-\frac{1}{4\pi^2\Xi}\int d\widetilde{\mathbf{x}}d\widetilde{\mathbf{x}}^{\prime}\widetilde{\rho}_f(\widetilde{\mathbf{x}})
\widetilde{\rho}_f(\widetilde{\mathbf{x}}^{\prime})\biggl[\widetilde{v}(\vert\widetilde{\mathbf{x}}-\widetilde{\mathbf{x}}^{\prime}\vert)/2
   -\widetilde{v}(\vert\widetilde{\mathbf{x}}-\widetilde{\mathbf{x}}_0\vert)\biggr]\right).
 \label{eq1.13}
\end{align}
From equation \eqref{eq1.10} we see that in the SC limit $1/\Xi\rightarrow 0$ only the lowest order canonical
partition function or the single polymer partition $Q_1$, contributes the most to the grand partition function. 
In the SC regime the counterions crowd near the oppositely charged fixed charge distribution and
the interactions with the surface dominates over the other interactions like the inter-polymer and excluded volume interactions. 
Thus the system effectively behaves like a single polymer system with an external potential due to the surface. These 
two polymer interactions are however are very important in the next order corrections to the partition function.
The point particle case is obtained by using $Q_1[i\phi] = \int d\widetilde{\mathbf{x}}e^{-i\phi(\widetilde{\mathbf{x}})}$. 

 We define the rescaled density distribution as $\widetilde{\rho}(\widetilde{\mathbf{x}})=2\pi\Xi\delta\ln Z/\delta h(\widetilde{\mathbf{x}})$.
 Similarly the density has an expansion in $1/\Xi$ as 
\begin{align}
 \widetilde{\rho}(\widetilde{\mathbf{x}})=\widetilde{\rho}_0(\widetilde{\mathbf{x}})+\frac{1}{\Xi}\widetilde{\rho}_1(\widetilde{\mathbf{x}}) +
 \mathcal{O}(1/\Xi^2).
 \label{eq1.14}
\end{align}
The zeroth order term of the density is derived from equations \eqref{eq1.11} and \eqref{eq1.12} of the partition function   
 \begin{align}
 \widetilde{\rho}_0(\widetilde{\mathbf{x}}) & = \frac{\delta Q_1[ h - \widetilde{u}]}{\delta h(\widetilde{\mathbf{x}})} \biggl\vert_{h=0}\nonumber \\
 & = \rho^{(1)}(\widetilde{\mathbf{x}};\widetilde{u}),
 \label{eq1.15}
 \end{align}
where $\rho^{(1)}(\widetilde{\mathbf{x}};\phi) = \delta Q_1[\phi]/\delta\phi(\widetilde{\mathbf{x}})$ is the single polymer density
 operator in an external field $\phi$  \cite{fredrickson2006equilibrium}. 
This is a main result of this paper. In the SC regime the counterions are located so close to the surface that the dominant contribution 
to the density is primarily due to the external potential due to the charged surface. In terms of the Wigner liquid picture
\cite{dobrynin2000adsorption,dobrynin2001adsorption,shklovskii1999wigner,shklovskii1999screening,perel1999screening,RevModPhys.74.329}, 
the lowest order description would 
be valid if the length of the polymers is lower than the size of the Wigner cell. For polymers which are slightly displaced from the 
nearest layer to the surface the next order term in the density becomes important. The first order term of the density
distribution function has the form
	\begin{align}
	\widetilde{\rho}_1(\widetilde{\mathbf{x}}) & = \frac{\delta (Q_2[ h - \widetilde{u},\widetilde{v}]-Q_1^2[ h - \widetilde{u}])}{\delta
		h(\widetilde{\mathbf{x}})} \biggl\vert_{h=0}.
	\label{eq1.16}
	\end{align}

\section{salt free case}
\label{Sec2}
We first consider the case when there is no salt in the system. In this case the electrostatic interaction potential in equations 
\eqref{eq1.1}-\eqref{eq1.16} is the Coulomb potential,
$\widetilde{v}(\vert\widetilde{\mathbf{x}} -\widetilde{\mathbf{x}}^{\prime}\vert) = 1/\vert\widetilde{\mathbf{x}}-\widetilde{\mathbf{x}}
^{\prime}\vert$. We consider the confinement of the GPEs by impenetrable charged walls. In the rest of the discussions we 
confine ourselves to very strong coupling regime. Thus our description would be based on the zeroth order calculations. As seen 
in the above Section the system effectively behaves like a single particle system under the external potential of 
the oppositely charged wall. We discuss how to calculate the density profiles of
GPEs from the formalism developed in the above Section and use them to qualitatively explain the adsorption and depletion phenomena.
The adsorption and depletion would be treated in a systematic way in the next section when salt is introduced.

\subsection{One charged wall}
\label{Sec2A}
We consider a system of GPEs confined in the upper $z$ plane by a charged plate located at $z = 0$. The rescaled charge distribution of the plate is 
$\widetilde{\rho_f}(\widetilde{\mathbf{x}}) = \delta(\widetilde{z})$. The wall potential from equation \eqref{eq1.8} is \cite{Netz}
\begin{equation}
 \widetilde{u}(\widetilde{z}) = \widetilde{z},
 \label{eq2A.1}
\end{equation}
where the reference point $\widetilde{\mathbf{x}_0}$ in the equation \eqref{eq1.8} is chosen to lie on the charged wall.

\begin{figure*}[h]
\centering
\includegraphics[scale = 0.5]{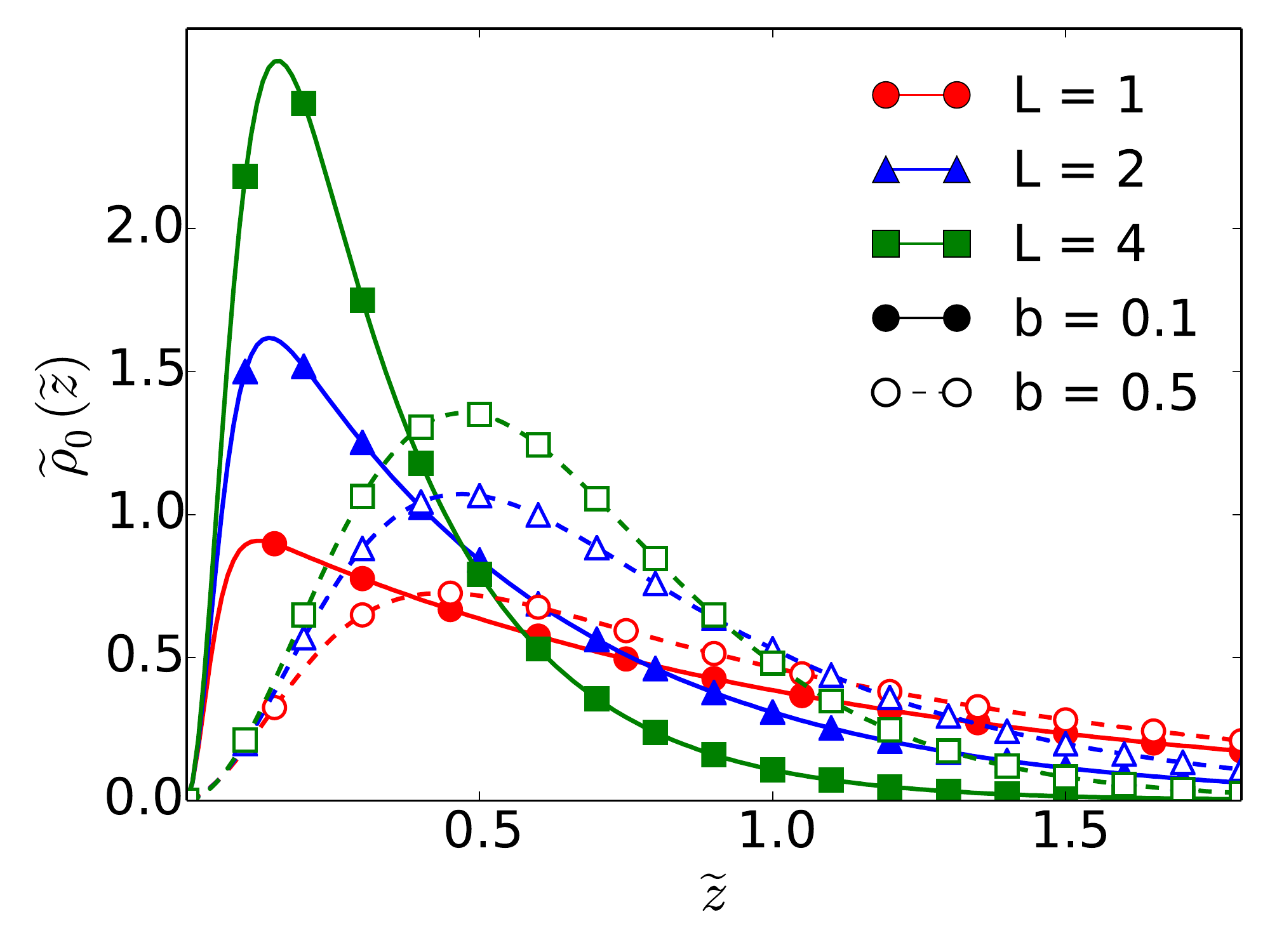}
\caption{Salt free polymer density confined in the upper $\widetilde{z} > 0$ half plane by a charged wall at $\widetilde{z} = 0$
 for different polymer lengths $L$ and monomer lengths $b$. The solid (filled) curves represent monomer length $b = 0.1$ and dashed (unfilled) curves 
 represent monomer length $b = 0.5$.}
 \label{Fig1}
\end{figure*}

In the SC regime, equation \eqref{eq1.15} says that the zeroth order term of the density distribution is the single polymer 
density of GPEs in an external potential of the wall potential
$\widetilde{u}(\widetilde{z})$. We use a procedure developed by Edwards 
\cite{doi1988theory,edwards1965statistical} to calculate the single polymer density by partial differential equations.
The zeroth order density distribution is then
\begin{align}
 \widetilde{\rho}_0(\widetilde{z})& = \Lambda_0\widetilde{\Omega}(\widetilde{z})\int_0^Ndsq(\widetilde{z},N-s;\widetilde{u})
 q(\widetilde{z},s;\widetilde{u}),
 \label{eq2A.2}
\end{align}
where the function $q(\widetilde{z},s)$ satisfies the following differential equation
\begin{align}
\left(\frac{\partial}{\partial N} - \frac{b^2}{6}\frac{\partial^2}{\partial\widetilde{z}^2} + \widetilde{u}(\widetilde{z})\right)q(\widetilde{z};N) = 0.
 \label{eq2A.3}
\end{align}
The function $q(\widetilde{z};N)$ satisfies the constraint $q(\widetilde{z};0) = 1$. Since the walls are impenetrable the density of the GPEs
and hence $q(\widetilde{z};N)$ vanishes on the wall surface, $q(0;N) = 0$.
The ion confinement function $\widetilde{\Omega}(\widetilde{z})$ in equation \eqref{eq2A.2} for a single wall confinement is
\begin{align}
 \widetilde{\Omega}(\widetilde{z}) & = 1 \hspace{2mm}\text{for } 0 < \widetilde{z}, \nonumber\\
 &  = 0 \hspace{2mm}\text{for }  \widetilde{z} < 0.
 \label{eq2A.4}
\end{align}
$\Lambda_0$ in equation \eqref{eq2A.2} is obtained from the charge neutrality condition \cite{Netz}
\begin{equation}
\int d\widetilde{z} \widetilde{\rho}_0(\widetilde{z}) = 1,
\label{eq2A.5}
\end{equation}
from which we obtain the zeroth order density as 
\begin{align}
 \widetilde{\rho}_0(\widetilde{z})& = \frac{\widetilde{\Omega}(\widetilde{z}) \int_0^Ndsq(\widetilde{z},N-s;\widetilde{u})q(\widetilde{z},s;\widetilde{u})}
 {\int d\widetilde{z}\widetilde{\Omega}(\widetilde{z})\int_0^Ndsq(\widetilde{z},N-s;\widetilde{u})q(\widetilde{z},s;\widetilde{u})}
 \label{eq2A.6}
\end{align}

The density profile is that of adsorption when the density of the PEs in the vicinity of the wall is higher than their bulk density. 
Figure \ref{Fig1} shows that the density profile is that of adsorption for lengths of the GPEs $L = 1$, $2$ and $4$ for monomer lengths
$b = 0.1$ and $0.5$. Since the longer chains
have higher charge, they have higher electrostatic attractions with the oppositely charged wall. Hence they are more strongly adsorbed into the wall
and their density distribution falls sharply compared to the shorter chains. GPEs with longer monomer lengths $b$, but having the same number 
of monomers $L$ and charge, have lower surface charge
density. Their attractions with the wall are weaker and they distribute further from the wall. Also the orientational degrees of freedom of chains 
with the same number of monomer but with larger monomer size is severely restricted near the wall. Therefore it is more entropically favorable for them
to be situated farther from the wall. We look at this adsorption behavior of GPEs in detail and introduce a quantity called polymer fraction to quantify the adsorption
phenomena in the next Section.

\begin{figure*}[h]
\centering
\includegraphics[scale = 0.5]{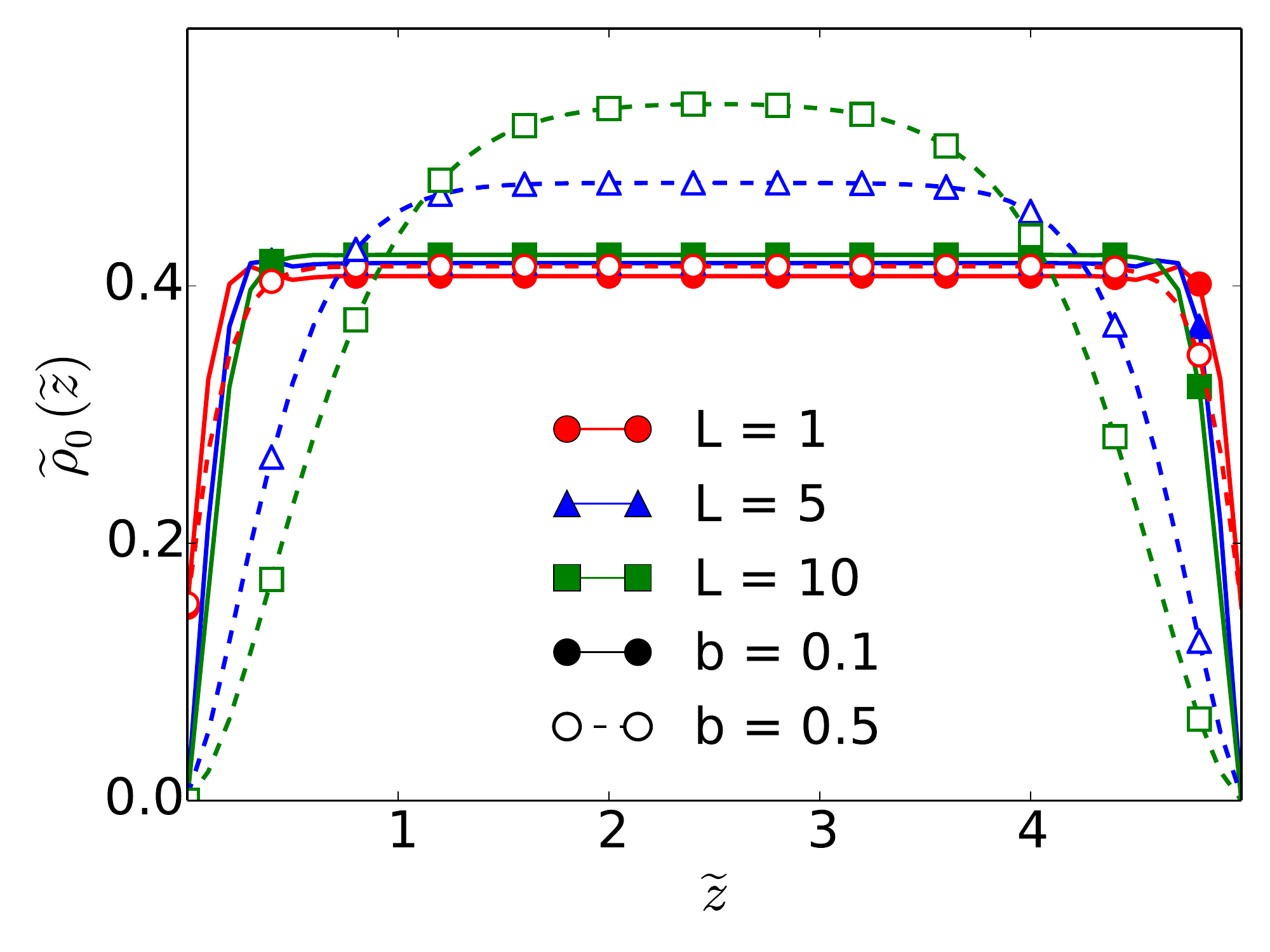}
\caption{Salt free polymer density confined between two charged walls with an inter-wall separation of $\widetilde{d}$
 for different polymer lengths $L$ and monomer lengths $b$. The solid (filled) curves represent monomer length $b = 0.1$ and dashed (unfilled) curves 
 represent monomer length $b = 0.5$.}
\label{Fig2}
\end{figure*}

\subsection{Two charged walls}
\label{Sec2B}
For GPEs confined between two charged plates located at $\widetilde{z} = 0$ and $\widetilde{z}=\widetilde{d}$ respectively,
the rescaled wall potential in this case is \cite{Netz}
\begin{equation}
 \widetilde{u}(\widetilde{z}) = 0,
 \label{eq2B.1}
\end{equation}
where the reference point $\widetilde{\mathbf{x}_0}$ in the equation \eqref{eq1.8} is chosen to lie at the midplane between the two walls.

The zeroth order density profile is given by the same equation \eqref{eq2A.2} except $q$ now satisfies the following differential equation
\begin{align}
\left(\frac{\partial}{\partial N} - \frac{b^2}{6}\frac{\partial^2}{\partial\widetilde{z}^2} \right)q(\widetilde{z};N) = 0,
 \label{eq2B.2}
 \end{align}
with impenetrable boundary conditions at the walls $q(0;N) = 0$ and $q(\widetilde{d};N) = 0$. Also the ion confinement function
$\widetilde{\Omega}(\widetilde{z})$ is modified to
\begin{align}
 \widetilde{\Omega}(\widetilde{z}) & = 1 \hspace{2mm}\text{for } 0 < \widetilde{z} < \widetilde{d}, \nonumber\\
 &  = 0 \hspace{2mm}\text{otherwise }.
\label{eq2B.3}
\end{align}

Figure \ref{Fig2} shows that the density profile of the GPEs is that of depletion for all lengths of the chains. Due to weaker attractions with 
the wall the chains with longer monomers having smaller charge density move towards the midplane between the two walls. Also the chains with 
smaller monomers have strong repulsions which makes it more energetically favorable for them to be spread nearly uniformly between the walls.

\section{Salt}
\label{Sec3}
To study the effect of the salt on our system, we use a screened potential
\begin{equation}
\widetilde{v}(\vert\widetilde{\mathbf{x}}-\widetilde{\mathbf{x}}^{\prime}\vert)
= \exp\left(-\kappa\vert\widetilde{\mathbf{x}}-\widetilde{\mathbf{x}}^{\prime}\vert\right)/
\vert\widetilde{\mathbf{x}}-\widetilde{\mathbf{x}}^{\prime}\vert,
 \label{eq3.0}
\end{equation}
that depends on the salt concentration $c$ through the 
screening length. The screening length $\kappa^{-1} = (8\pi l_Bc)^{-1/2}$ is made dimensionless by scaling with the Guoy-Chapman-like length $\mu$ 
and can be expressed as a function of the dimensionless salt density $\widetilde{c} = c\mu^{3}$ by
\begin{equation}
 \widetilde{\kappa} = \kappa\mu = (8\pi \Xi \widetilde{c})^{1/2}.
 \label{eq3.1}
\end{equation}

\subsection{One wall}
\label{Sec3A}
Plugging equation \eqref{eq3.0} in equation \eqref{eq1.8} and by choosing the reference point on the wall $\widetilde{\mathbf{x}}_0 = 0$,
the wall potential becomes
\begin{align}
 \widetilde{u}(\widetilde{z}) 
 & = -\int_0^{\infty}\frac{\exp(-\widetilde{\kappa}\sqrt{\rho^2+\widetilde{z}^2})}{\sqrt{\rho^2+\widetilde{z}^2}}\rho d\rho + 
     \int_0^{\infty}\exp(-\widetilde{\kappa}\rho) d\rho \nonumber \\
 & = \frac{1}{\kappa}\left(1 - \exp(-\widetilde{\kappa}\widetilde{z})\right).
 \label{eq3A.1}
\end{align}
When there is no screening $\widetilde{\kappa} = 0$, we recover the wall potential $\widetilde{u}(\widetilde{z})$ in equation 
\eqref{eq2A.1}.

\begin{figure*}[h]
\centering
\subfloat[]{
\includegraphics[scale = 0.4]{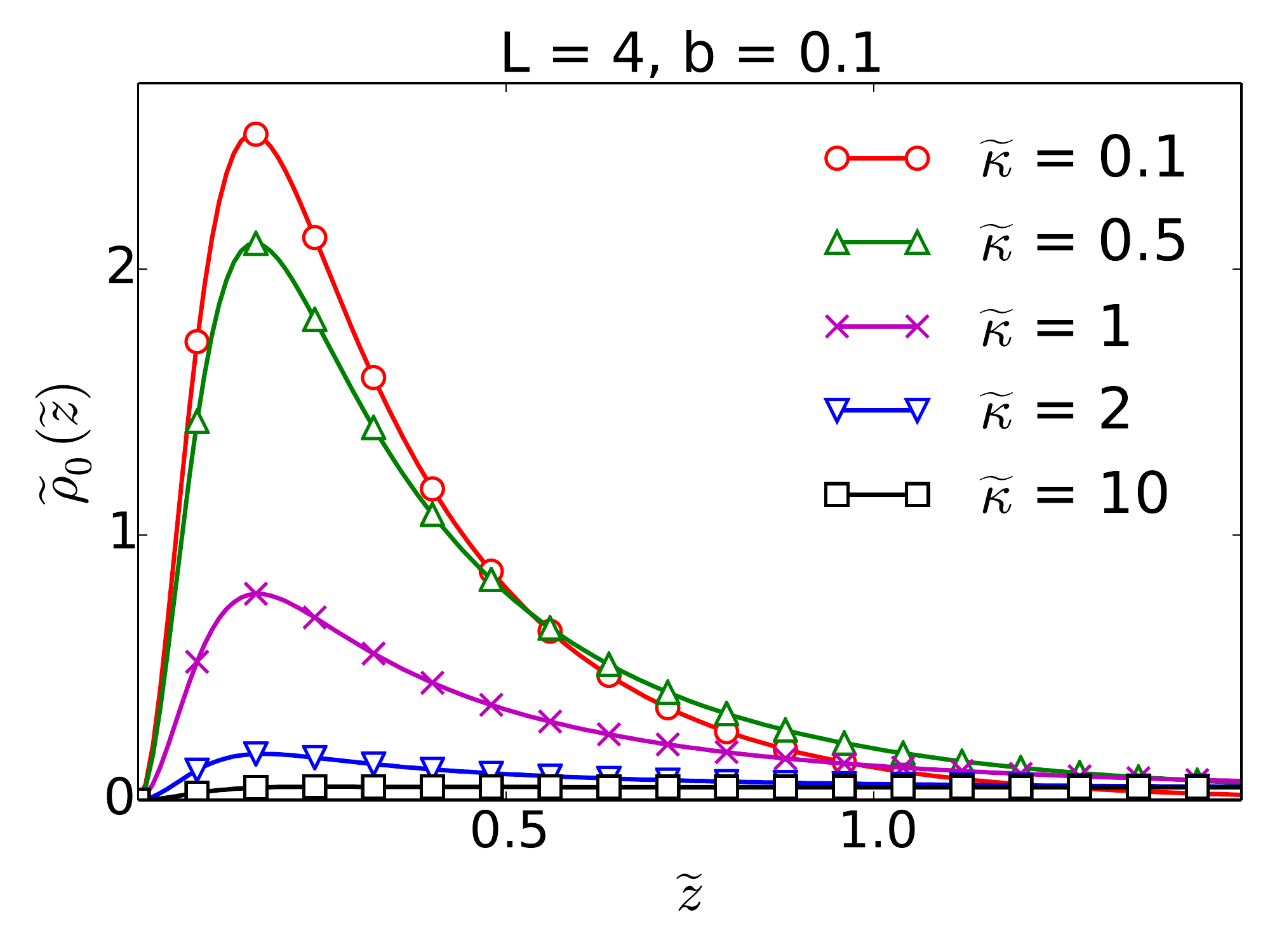}}
\subfloat[]{
\includegraphics[scale = 0.4]{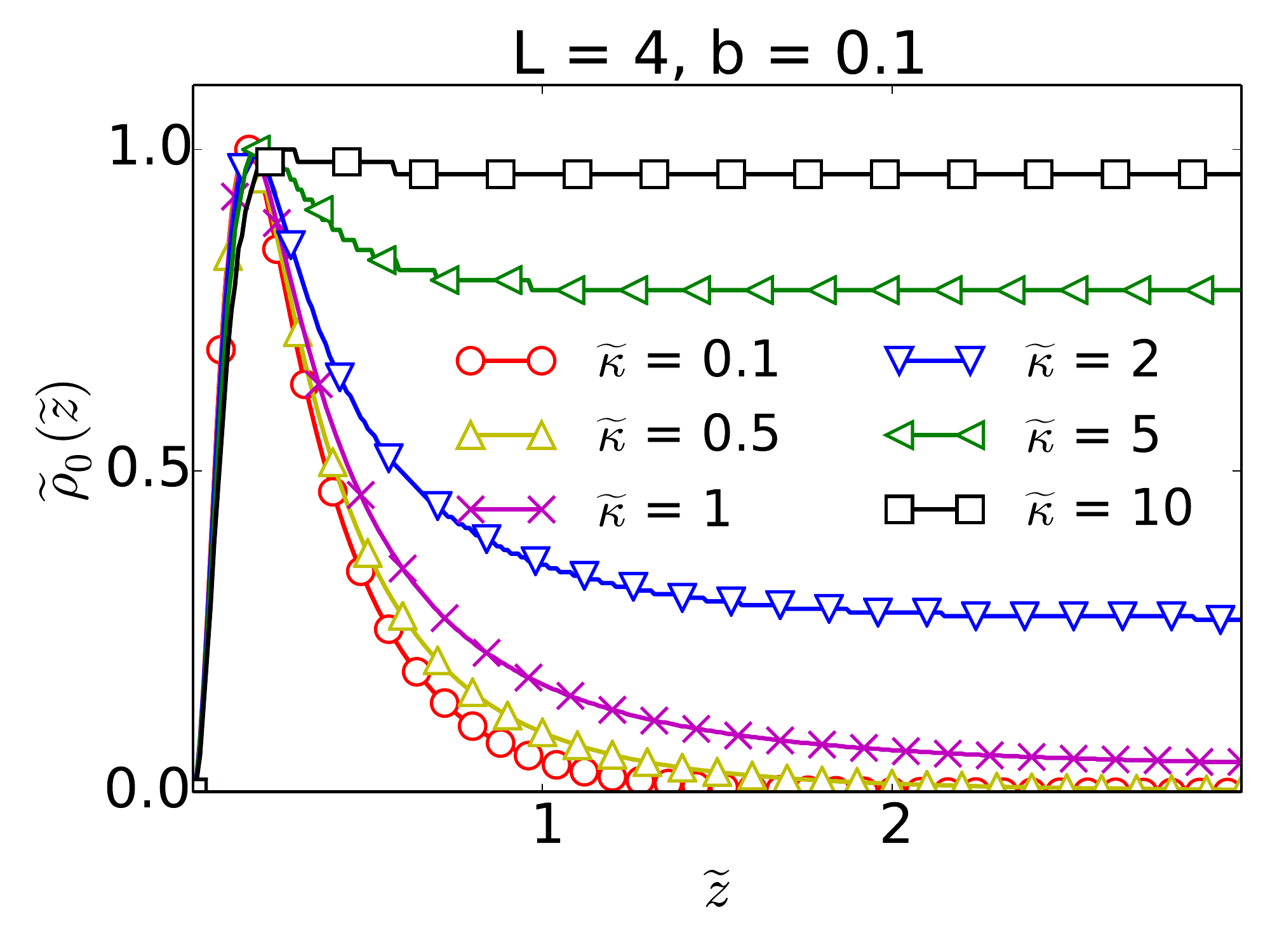}}
\caption{\label{Fig3} (a) Polymer density near a charged wall for various inverse screening lengths $\kappa$. (b) Polymer densities
in (a) scaled by their corresponding maximum values.}
\label{Fig3}
\end{figure*}

\begin{figure*}[h]
\centering
\subfloat[]{
\includegraphics[scale = 0.4]{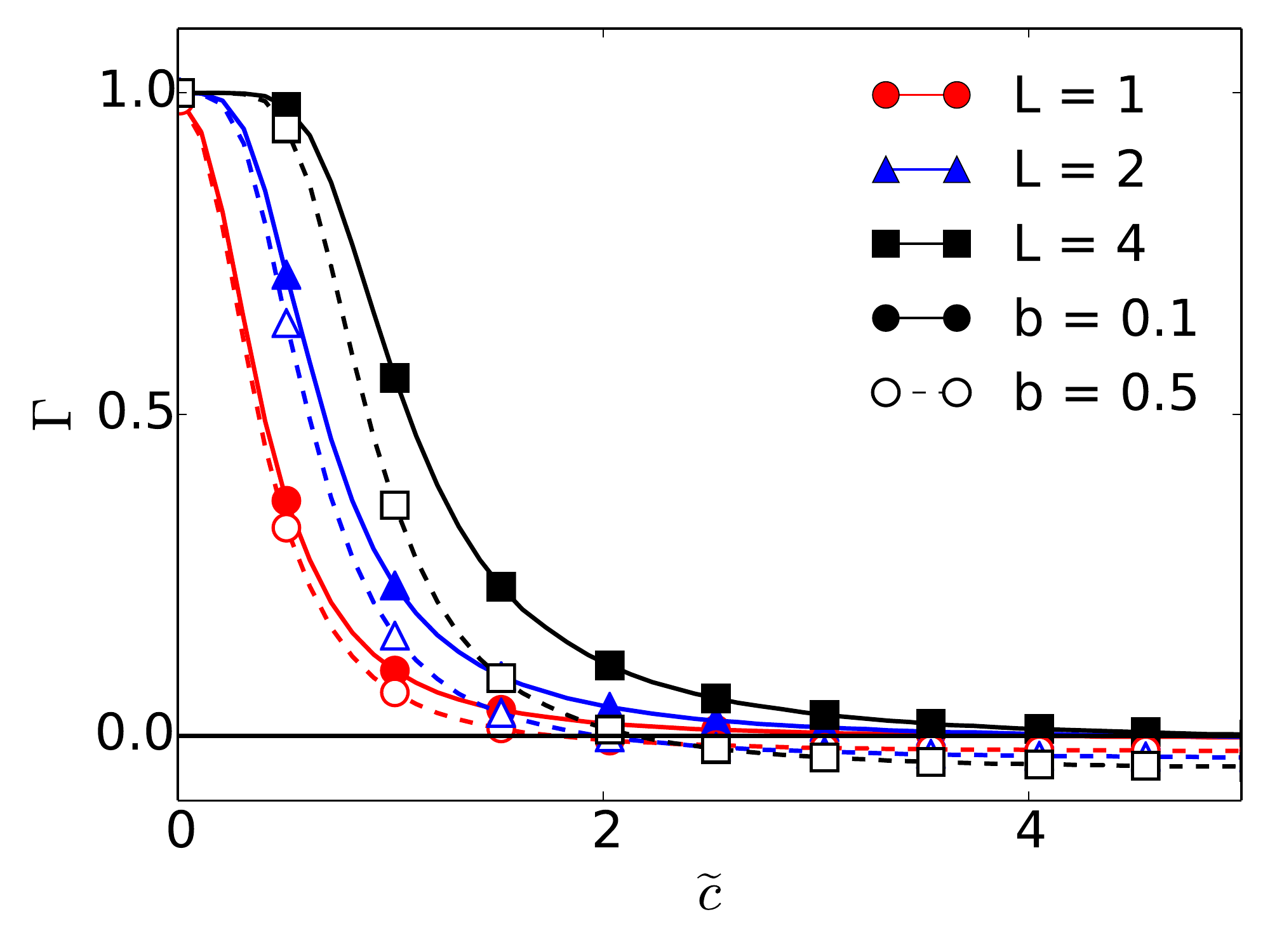}}
\subfloat[]{
\includegraphics[scale = 0.4]{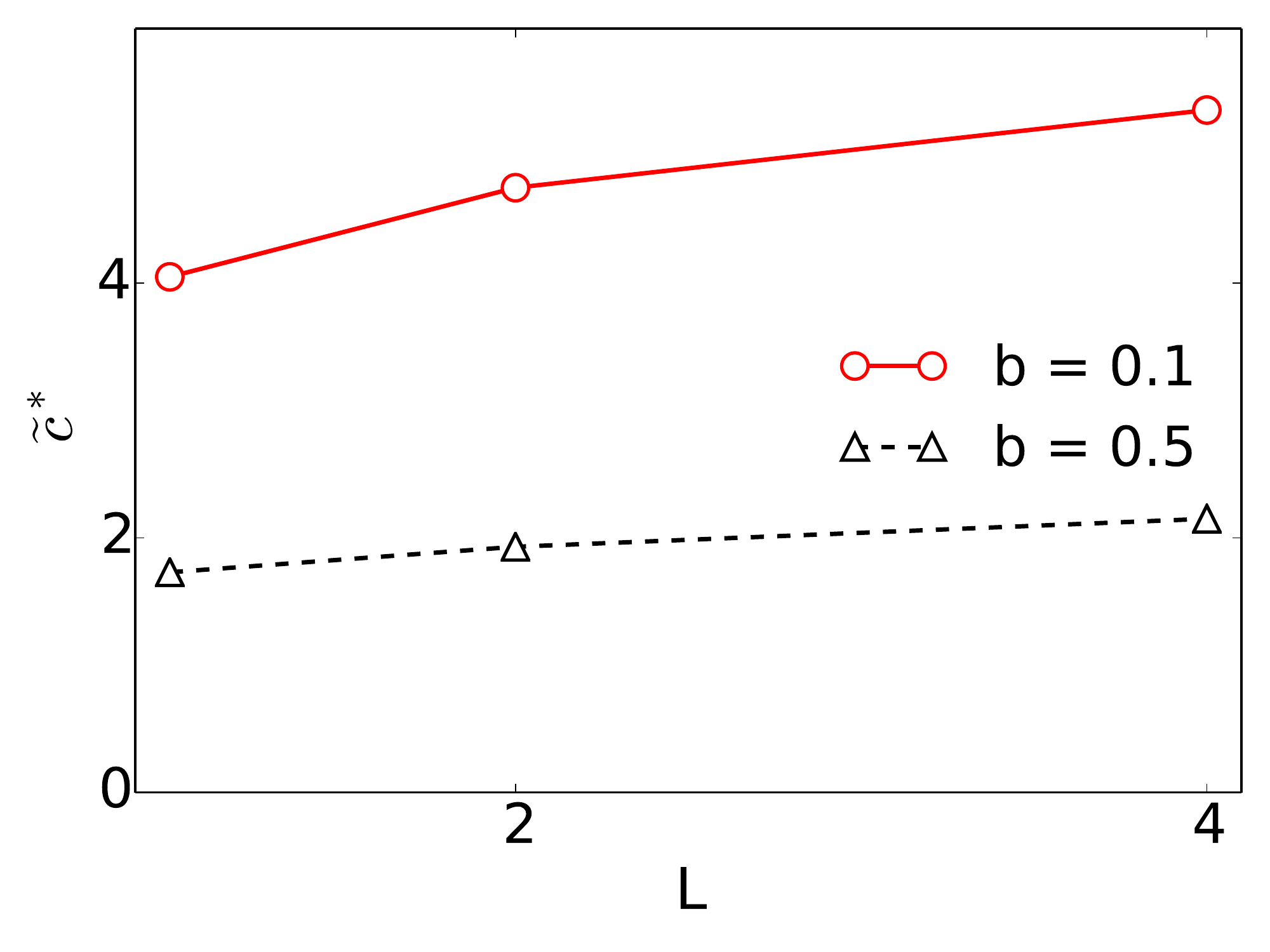}}
\caption{(a) The fraction of polymers adsorbed into the wall $\Gamma$ vs the salt concentrations $\widetilde{c}$ for various number of monomers $L$
and monomer lengths $b$. At low salt concentrations $\Gamma > 0$ showing adsorption while at large salt concentration $\Gamma < 0$ which signals
depletion. The solid (filled) curves represent monomer length $b = 0.1$ and dashed (unfilled) curves 
 represent monomer length $b = 0.5$. (b) The critical salt concentration $\widetilde{c}^{\ast}$ for adsorption-depletion transition vs the polymer lengths $L$ and monomer lengths
$b$. }
\label{Fig4}
\end{figure*}

Figure \ref{Fig3} shows the density of GPEs in the presence of salt. Increasing $\kappa$ decreases the electrostatic correlations
between the wall and the counterions. This decreases the distribution of polymers near the wall. Figure \ref{Fig3}-(b) depicts 
the density profile as in Figure \ref{Fig3}-(a) but each curve is scaled with its maximum value. On increasing the screening
the bulk polymer develops a non-zero value. The plot also shows the transition 
of the density distribution from adsorption, when the density near the wall is much higher than the bulk density, to depletion, 
when the density is almost flat. At low screening the strong attractions with the walls cause almost all
the polymers to be adsorbed into the wall and the bulk polymer concentration is zero. On increasing the screening the attractions decreases
and more and more GPEs leave the wall and move to the bulk causing depletion. We can make a more accurate estimate of the transition by measuring the 
total amount of polyelectrolytes in the adsorbed layer \cite{shafir2003adsorption} by
\begin{equation}
 \Gamma = \int_0^{\infty}d\widetilde{z}\left(\widetilde{\rho}_0(\widetilde{z})-\widetilde{\rho}_0^b\right),
\label{eq3A.2}
\end{equation}
where $\widetilde{\rho}_0^b = \widetilde{\rho}_0(\widetilde{z}=\infty)$ is the bulk density of the GPEs. In the adsorption region
$\Gamma$ is positive while in the depletion region $\Gamma $ is negative. Thus $\Gamma = 0$ would signify the transition from 
adsorption to depletion as shown in Figure \ref{Fig4}-(a) by varying the salt concentration $\widetilde{c}$. $\Gamma = 1$ at 
low salt concentrations implies almost all the polymers are adsorbed into the wall. On increasing the salt, the entropic forces takes over
the electrostatic correlations and the polymers spread away from the wall. Figure \ref{Fig4}-(b) plots the critical salt concentration
$\widetilde{c}^{\ast}$ when the polymer density profile changes from adsorption to depletion, $\Gamma = 0$.  
Longer chains have stronger electrostatic attractions with the wall and hence their transition from the adsorption to depletion 
transition occurs at a higher salt concentration as seen from the same Figure. Chains with shorter monomer lengths have higher charge densities
and hence stronger electrostatic correlations and need higher salt screening to overcome them. As noted earlier chains with longer 
monomers have much less orientational freedom closer to the wall, hence it is easier to salt them out.

\subsection{Two wall}
\label{Sec3B}

\begin{figure*}[h]
\centering
\includegraphics[scale = 0.5]{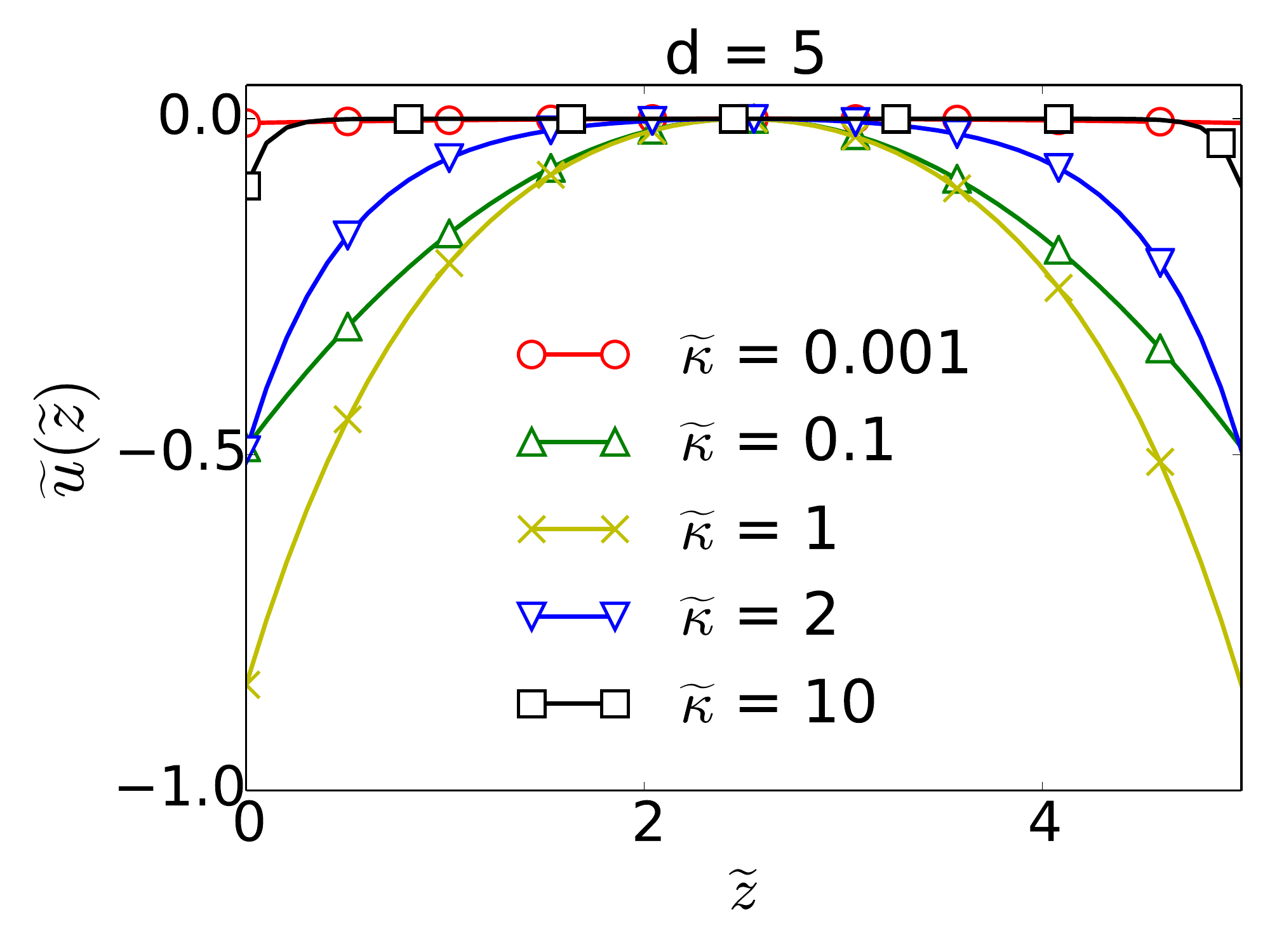}
\caption{ The dependence of the wall potential $\widetilde{u}(\widetilde{z})$ on the inverse screening length $\kappa$ with the inter-wall distance
is $\widetilde{d} = 5$.}
\label{Fig5}
\end{figure*}

\begin{figure*}[h]
        \centering
           \subfloat{%
              \includegraphics[height=6.2cm]{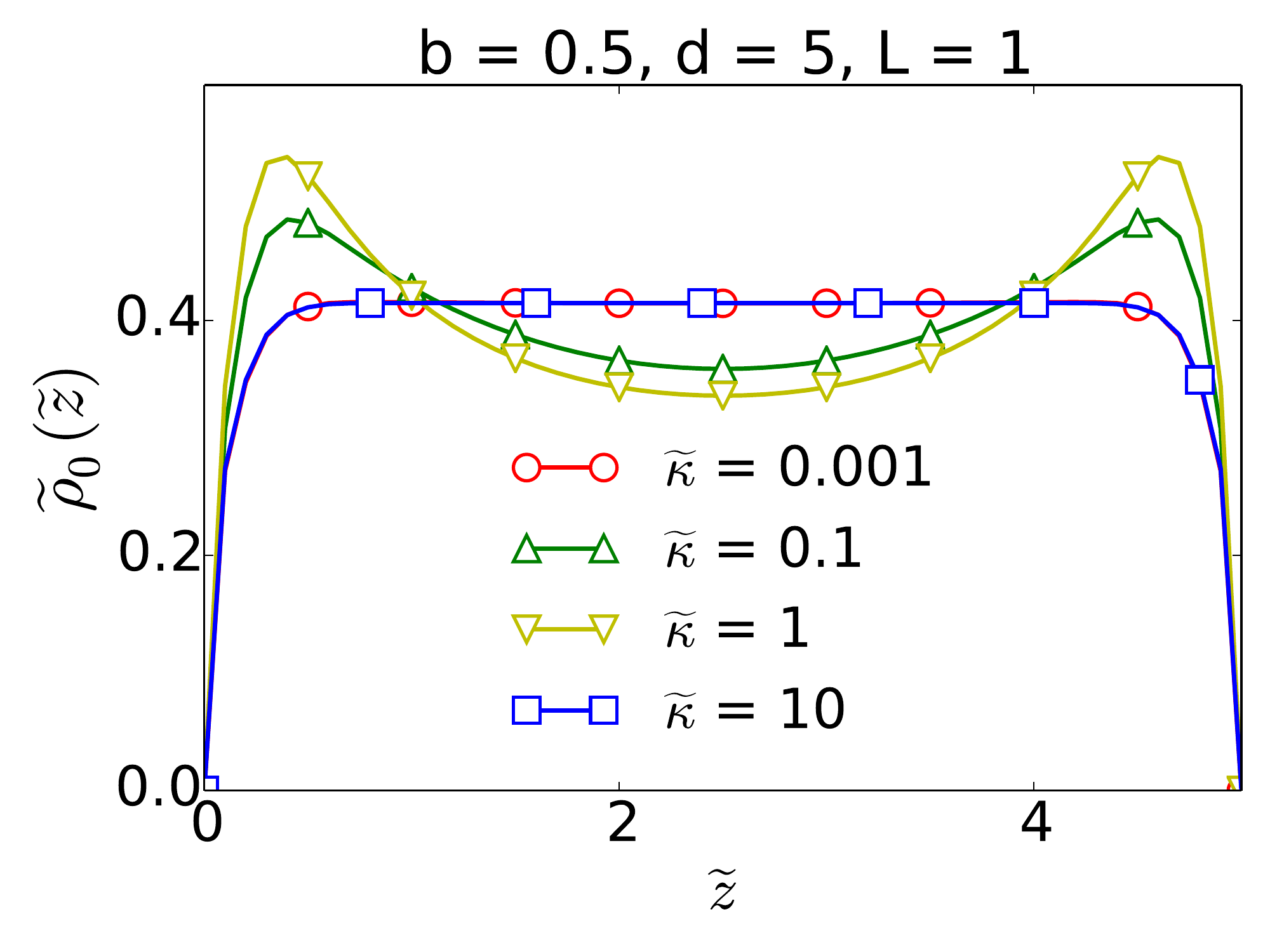}%%
           }
           \subfloat{%
              \includegraphics[height=6.2cm]{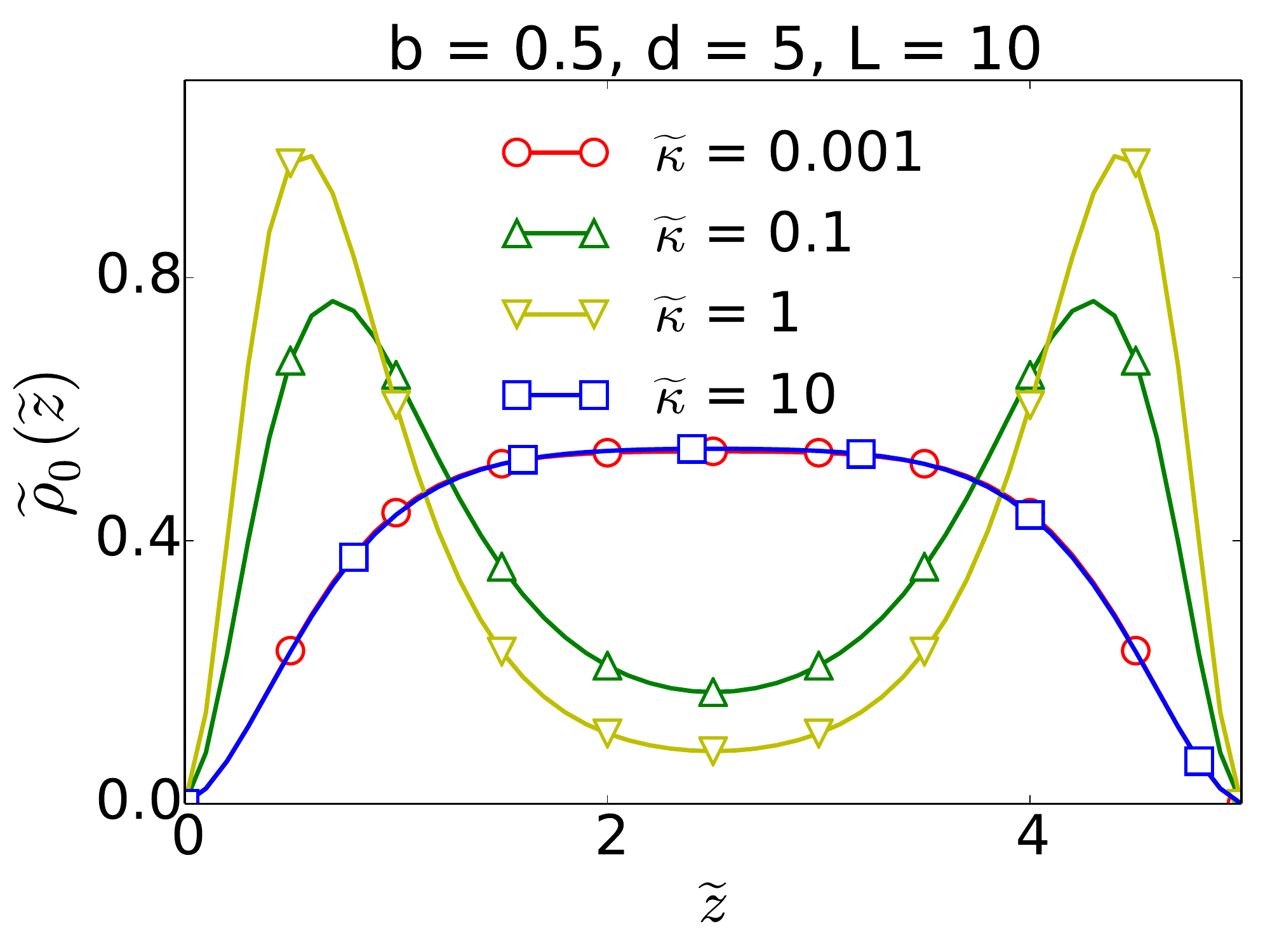}%
           }
           \caption{Polymer densities vs the inverse screening lengths $\kappa$ with inter-wall separation of $\widetilde{d} = 5$ for 
           polymer lengths (a) $L = 1$ and (b) $L = 10$.}
           \label{Fig6}
 \end{figure*}
 
\begin{figure*}[h]
\centering
\subfloat[]{
\includegraphics[scale = 0.4]{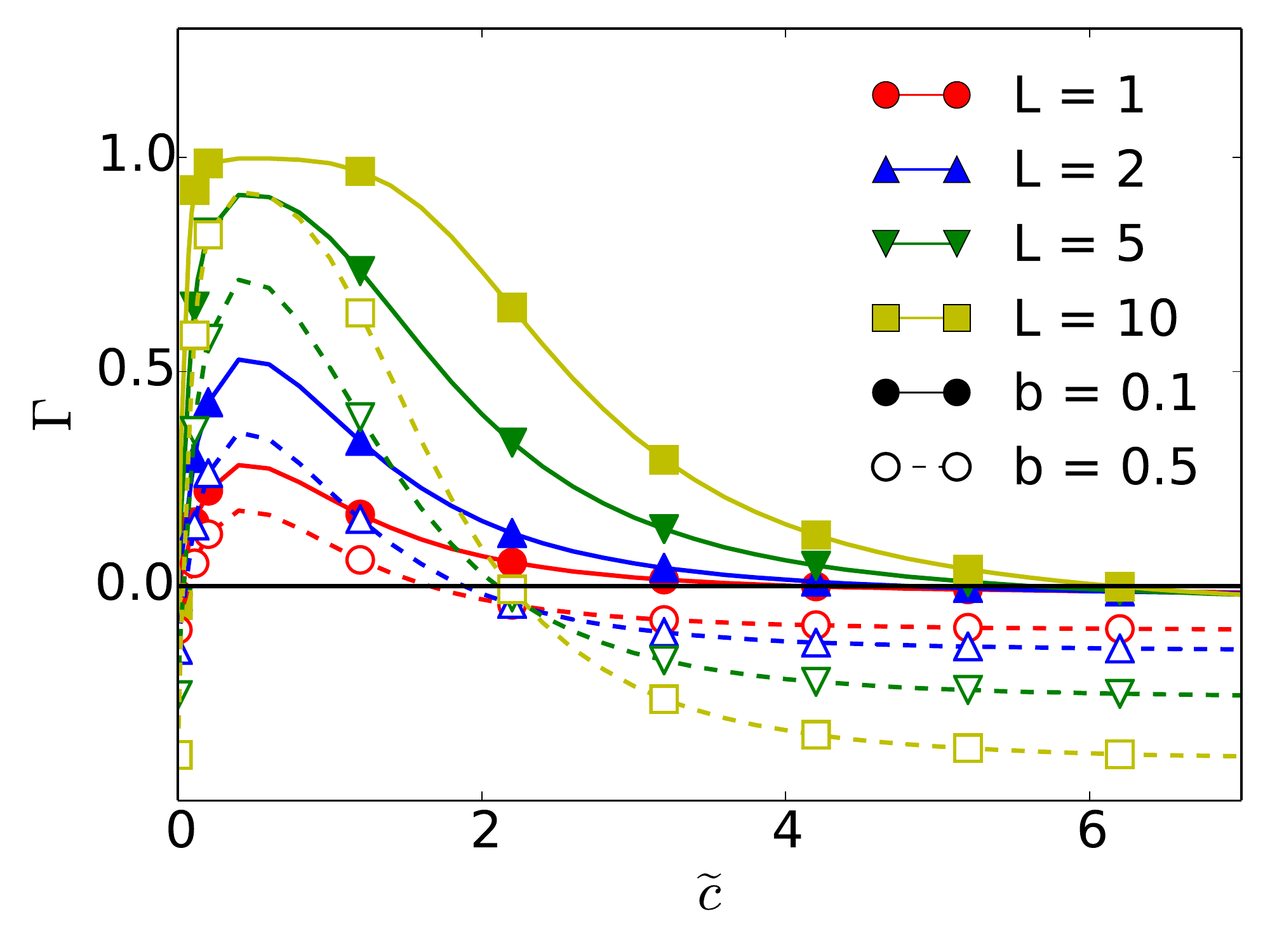}}
\subfloat[]{
\includegraphics[scale = 0.4]{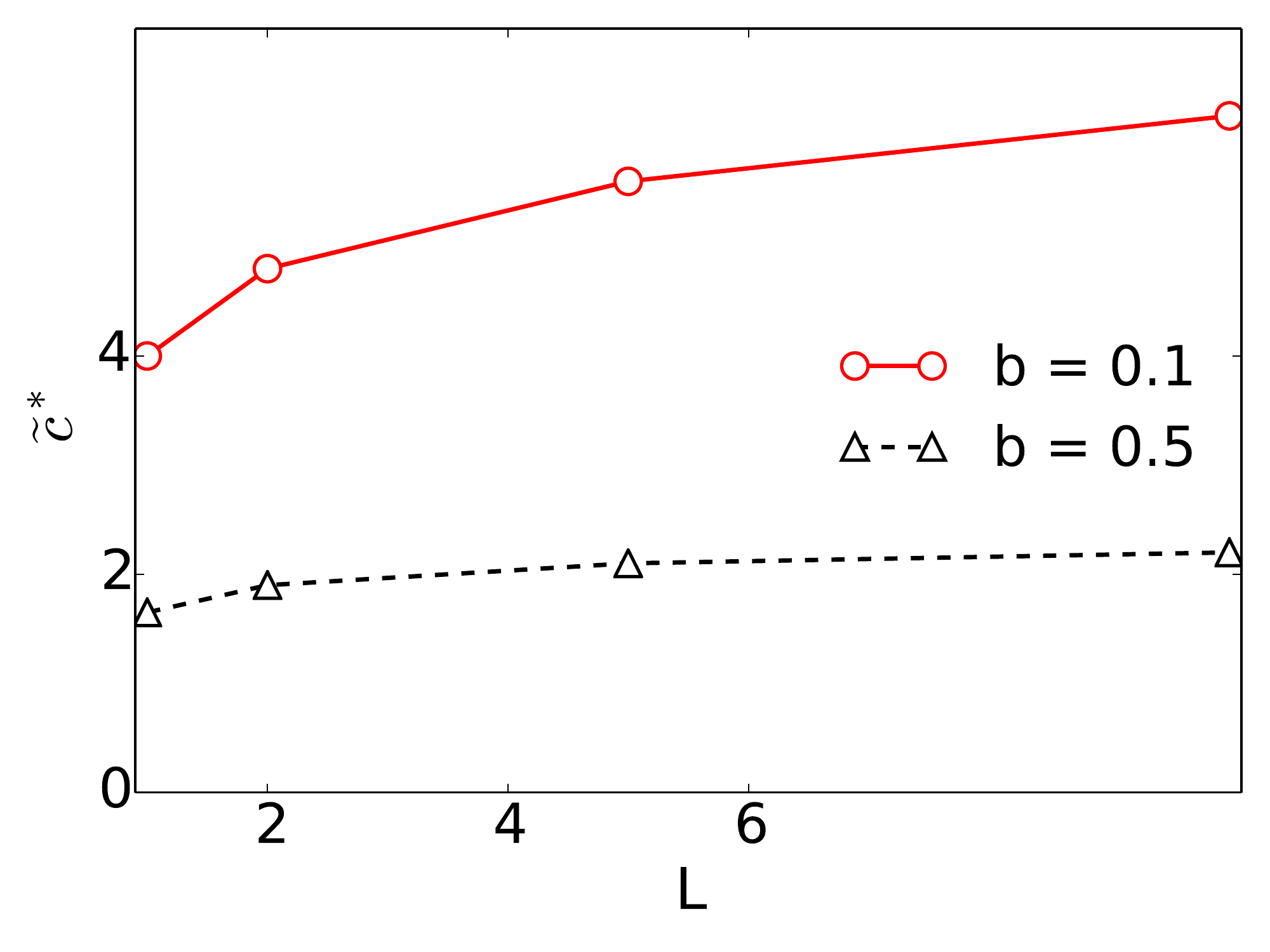}}
\caption{(a) The polymer adsorption fraction showing the depletion-adsorption-depletion transitions on increasing the salt concentrations $\widetilde{c}$.
The solid (filled) curves represent monomer length $b = 0.1$ and dashed (unfilled) curves represent monomer length $b = 0.5$.
(b) The dependence of the critical salt concentration on the polymer lengths $L$.}
\label{Fig7}
\end{figure*}

When there are two walls, the wall potential is obtained from equation \eqref{eq3A.1} for the one wall case or from equations \eqref{eq1.8} and
\eqref{eq3.0} with the reference point $\widetilde{\mathbf{x}}_0$ at the midplane between the two walls
\begin{equation}
 \widetilde{u}(\widetilde{z}) = \frac{1}{\kappa}\left(2\exp(-\kappa\widetilde{d}/2)-\exp(-\kappa\widetilde{z})-\exp(-\kappa\vert\widetilde{d}-
 \widetilde{z}\vert)\right).
 \label{eq3B.1}
\end{equation}
We plot the wall potential $\widetilde{u}(\widetilde{z})$ in Figure \ref{Fig5} for various values of the screening parameter $\widetilde{\kappa}$
for the inter-wall separation of $\widetilde{d} = 5$. When the salt concentration is zero or $\widetilde{\kappa} \rightarrow 0$,
$\widetilde{u}(\widetilde{z}) = 0$ as in equation \eqref{eq2B.1}. When salt concentration $\widetilde{c}$ and hence $\widetilde{\kappa}$ is large,
again we see from the Figure that $\widetilde{u}(\widetilde{z}) \approx 0$ especially when the walls are further apart. When there is no screening 
the electric fields of the walls are equal and opposite and cancel each other because of the planar nature of the walls. In the other 
limit of very high screening the potential is extremely short ranged and the potential is zero everywhere except very close to the walls.
Thus effectively in the two limits the wall potential has the same form. As a result the density profile of GPEs are almost similar at low and 
high values of the screening parameter in Figure \ref{Fig6} irrespective of the lengths of the polymers. Since longer chains have 
higher charges they are more clustered towards the wall as shown in the Figure. In Figure \ref{Fig7}-(a) the polymer adsorption fraction shows 
a transition of the density profile from depletion in the salt free regime to adsorption due to the variation
of the wall potential from the zero electric
field Coulomb to screened Yukawa regime. On further increasing the salt, the system again undergoes a depletion transition. In
Figure \ref{Fig7}-(b) the critical salt concentration for the adsorption-depletion transition is shown for different lengths of the
polymers. It shows that the longer chains needs higher salt concentration to screen out their electrostatic correlations.

\section{Conclusions and discussions}
\label{Sec4}
We have presented a theoretical formalism to describe highly charged Gaussian polymer in presence of a fixed charged distribution.
Using polymer field theory we have developed the formalism closely following the strong coupling theory by Netz and his coworkers
\cite{Netz,moreira2001binding}. We considered the confinement of the GPEs by oppositely charged impenetrable walls in the presence
and absence of salt. In the presence of salt we consider the confinement by one and two charged walls and 
study the dependence of the density profile on the geometry of the chains and the salt concentration. At low salt
concentration and for confinement by a single wall, the polymers are strongly adsorbed into the wall especially the longer chains since they have stronger electrostatic 
attractions with the wall. On further increasing the salt concentration the chains de-adsorb from the wall and their density
distribution shows a depletion transition. The dependence of the critical salt concentration at the adsorption-depletion transition
on the geometry of the polymers is discussed. For the confinement between two walls, the wall potential is zero when it is salt-free
and almost zero when there is high salt concentration. When there is no salt the electric fields due to the two plates are equal and opposite and they 
cancel each other. When there is large amount of salt in the solution the screening is strong and the potential is zero except 
very close to the walls. Thus in the two limits there are depletion and in the intermediate regime there is adsorption.

Our analysis is however valid for very strong coupling $\Xi >> 1$, because of which we consider only the zeroth order term of the density.
The single particle description is appropriate when the polymer is shorter then the Wigner cell.
This analysis would be true for the layer closest to the walls, but for the polymers located off the condensed layer the higher order
terms, which have the excluded volume and the two polymer electrostatic interaction effects, have to be taken into consideration.
The intermediate coupling regime which occurs in most real systems. Most biopolymers also have finite bending rigidity and semi-flexible
polymers not Gaussian polymers are a better description for them. 
Extension of this analysis to semiflexible polymers will be done in a subsequent paper.

\bibliography{bibliography}

\end{document}